\documentclass[conference]{IEEEtran}
\IEEEoverridecommandlockouts
 \usepackage{lipsum,graphicx,multicol}
\usepackage{color,soul}
\usepackage{mathtools}
\usepackage{graphicx}
\usepackage{color,soul}
\usepackage{setspace}
\usepackage{multirow}
\usepackage{tablefootnote}
\usepackage[norelsize, linesnumbered, ruled, lined, boxed, commentsnumbered]{algorithm2e}
\usepackage[normalem]{ulem}
\usepackage{comment}
\usepackage[font=small,labelfont=bf]{caption}
\usepackage{subcaption}
\usepackage{bm, amsmath, amssymb, amsthm}
\usepackage{amsfonts}
\usepackage {amssymb,amsmath,xcolor}
\usepackage{booktabs}
\usepackage{cite}
\usepackage{array}
\usepackage[acronym]{glossaries}
\usepackage{cases}
\usepackage{cleveref}

\def\BibTeX{{\rm B\kern-.05em{\sc i\kern-.025em b}\kern-.08em
    T\kern-.1667em\lower.7ex\hbox{E}\kern-.125emX}}
    
  \newtheorem{theorem}{Theorem}

\newtheorem{problem}{Problem}
\newtheorem{corollary}{Corollary}
\newtheorem{proposition}{Proposition}
\newtheorem{lemma}{Lemma}
\newtheorem{remark}{Remark}

\newcommand{\bieee}{\begin{IEEEeqnarray}{rCl}}
\newcommand{\eieee}{\end{IEEEeqnarray}}

\usepackage{amsthm,mdframed,calc}

\SetKwInput{KwInput}{Input}                
\SetKwInput{KwOutput}{Output}              

\begin{document}

\title{On Securing Berrut Approximated Coded Computing Through Discrete Cosine Transforms
}

\author{\IEEEauthorblockN{Rimpi Borah and J. Harshan}
\IEEEauthorblockA{Department of Electrical Engineering, Indian Institute of Technology Delhi, India}}

\maketitle

\begin{abstract}
Coded computing is a reliable and fault-tolerant mechanism for implementing large computing tasks over a distributed set of worker nodes. While a majority of coded computing frameworks address accurate computation of the target functions, they are restricted to computing multivariate polynomial functions. To generalize these computing platforms to non-polynomial target functions, Jahani-Nezhad and Maddah-Ali recently proposed Berrut Approximated Coded computing (BACC), which was proven fault-tolerant against stragglers albiet with tolerable approximation errors on the target functions. Despite these benefits, there is no formal study on the security of BACC against worker nodes which report erroneous computations. To fill this research gap, we use a coding-theoretic approach to propose Secure Berrut Approximated Coded Computing (SBACC), which is resilient to stragglers and also robust to the presence of such untrusted worker nodes. One of the highlights of SBACC is the new choice of evaluation points for distributed computation which makes the well-known Discrete Cosine Transform (DCT) codes amenable to error detection and correction. To validate the new choice of evaluation points, first, we derive bounds on the accuracy of SBACC in the absence of untrusted worker nodes. Subsequently, to handle the presence of untrusted worker nodes, we derive bounds on the accuracy of SBACC and show that interesting optimization problems can be formulated to study the trade-off between the error correcting capability of the DCT codes and the accuracy of the target computation.
\end{abstract}



\section{Introduction}

Coded computing is an efficient and fault-tolerant mechanism of distributing large-scale computational tasks over a set of parallel computing machines referred to as worker nodes. While traditional coded computing was designed for robustness against stragglers \cite{b1}, \cite{w1}, their recent variants strive to achieve robustness in the presence of untrusted worker nodes, which may be interested in either learning the underlying datasets or perturbing the final computation by returning erroneous answers \cite{b3}, \cite{b4}, \cite{b5}, \cite{b6}. Towards developing robust coded computing frameworks against such threats, various coding-theoretic solutions have been proposed  \cite{a1}, \cite{a2} depending on the algebraic nature of the underlying target functions, the privacy demands on the datasets, and the worst-case number of untrusted worker nodes in the network. 

\subsection{Background and Motivation}

Although robustness against untrusted workers is required for fault-tolerance, privacy and integrity features, henceforth, in this work, we focus on coded computing frameworks that are robust to worker nodes that are stragglers and worker nodes that return erroneous computations. First, we provide a brief background on a coded computing framework that have the above properties, and highlight their operational features. In such a coded computing setup, a master server intends to apply a target function $f: \mathbb{R}^{m \times n} \rightarrow \mathbb{R}^{m \times n}$ on each of its multiple datasets, denoted by $\{\mathbf{X}_{0}, \mathbf{X}_{2}.\ldots, \mathbf{X}_{K-1} ~|~ \mathbf{X}_{j} \in \mathbb{R}^{m \times n}\}$. To implement this task in a distributed manner, the master server constructs an arbitrary function $u(z) = \sum_{j = 0}^{K-1} \gamma_{j}(z)\mathbf{X}_{j}$ in the variable $z$ involving all its datasets and the interpolant functions $\{\gamma_{j}(z)\}$ such that $\gamma_{j}(\alpha_{i}) = 0$ for $j \neq i$ and $\gamma_{j}(\alpha_{i}) = 1$ for $j = i$ over the points $\{\alpha_{j} \in \mathbb{R}, 1 \leq j \leq K\}$. Subsequently, using the evaluation points $\{z_{i} \in \mathbb{R}, 1 \leq i \leq N\}$, the master server distributes $u(z_i)$ to the $i$-th worker node for $i \in [N-1]$. Then the worker nodes apply the same function $f(\cdot)$ on their evaluations, and return the set $\{f(u(z_{i}))\}$ to the master server. Finally, upon receiving the computations from the worker nodes, the master server reconstructs $f(u(z))$ through interpolation, and then recovers the functional evaluation of its datasets $\{f(\mathbf{X}_{j})\}$ by evaluating the reconstructed $f(u(z))$ at $\{\alpha_{j}\}$. Under this framework, a majority of prior contributions have addressed $f(\cdot)$ that are multivariate polynomials. As a consequence, if $u(z)$ is a finite degree polynomial, then $f(u(z))$ will also be a finite degree polynomial, thereby ensuring that only a subset of computations from the $N$ worker nodes is sufficient for accurate reconstruction of $f(u(z))$. Thus, in such methods, resilience against stragglers is handled due to the finite degree nature of $f(u(z))$. On the other hand, when dealing with worker nodes that report erroneous computations, we note that the computations returned from some of the worker nodes are noisy versions of $\{f(u(z_{i}))\}$. Therefore, if any of erroneous values have to be detected or corrected, it is clear that $\{f(u(z_{i}))\}$ must form a codeword of a suitable error correcting code with good error correcting capability. Thus, the choice of the evaluation points $\{z_{i}\}$ plays a crucial role in achieving resilience against erroneous computation in distributed computation. For instance, with reference to LCC \cite{b3} and ALCC \cite{b6},\cite{a1},\cite{a2} that provide these features, $u(z)$'s are chosen as finite degree Lagrange polynomials and $\{z_{j}\}$ are chosen as primitive elements in their fields, which together ensure error correction capability as well as resilience to stragglers. 

In contrast to \cite{b3} and \cite{b6}, if $f(\cdot)$ is non-polynomial in nature, then $f(u(z))$ may not have a polynomial representation, and as a consequence, accurate reconstruction of $f(u(z))$ may not be possible using a finite number of samples $\{f(u(z_{i}))\}$. Subsequently, the target computation on the datasets suffer from approximation errors. To address this problem, Tayyebeh Jahani-Nezhad and Mohammad Ali Maddah-Ali proposed Berrut Approximated Coded Computing (BACC) \cite{b7} that provides resilience against stragglers albeit with tolerable approximation errors. The crux of the idea was to use Berrut's rational interpolants for the role of $\{\gamma_{i}(z) | 1 \leq i\leq K\}$, Chebyshev points of the first kind for $\{\alpha_{j} | 1 \leq j\leq K\}$ and the  Chebyshev points of the second kind for $\{z_{i} | 1 \leq i \leq N\}$. As a consequence, it was shown that the Lebesque constants of the Chebyshev points are bounded, and moreover, the error term resulting from distributed computation is a decreasing function of the number of worker nodes. Due to these significant benefits, BACC has also received some attention \cite{a3, b8}.

Despite providing reasonable accuracy benefits for non-polynomial functions, robustness of BACC has not been studied against workers that return erroneous computations to the master server. While there are no contributions that explicitly address this topic, ApproxIFER \cite{a3}, remotely intersects this space in the context of designing prediction serving systems. In particular, \cite{a3} leverages BACC to develop prediction serving systems that are robust to workers which return erroneous computations. Although ApproxIFER uses an algorithmic approach to localize the adversarial workers, (i) it assumes prior knowledge of the number of adversaries without providing an explicit method to estimate the same, and (ii) the computations from the localized adversaries are discarded since the reconstruction process does not capitalize on any systematic codes to correct the errors. Thus, if the approach used in ApproxIFER is used for securing BACC, it will result in accuracy loss owing to reduced number of reconstruction points, especially when the number of adversarial workers is not small.



Based on the above discussion, to secure BACC, there is a need for designing coding-theoretic approaches so that the errors introduced by the adversarial workers can be corrected in the reconstruction process. Towards that direction, we observe that the generator matrix of the underlying linear code in BACC neither has Bose–Chaudhuri–Hocquenghem (BCH) type of structure, nor is amenable for minium-distance characterization using low-complexity methods. Thus, there is no evidence to show that BACC supports error correction mechanisms to provide robustness against adversarial workers.  

\subsection{Contributions}
To overcome the limitations in the design of BACC, we present a new approximate coded computing framework referred as Secure BACC (SBACC), for computing non-polynomial functions in a distributed manner. As the main departure from BACC \cite{b7}, we propose Chebyshev points of the first kind as the evaluation points when the master distributes the computing tasks to the worker nodes. As an immediate consequence of this proposal, we show that the computations returned by the worker nodes can be decomposed into two portions, wherein, the first part is a codeword of a Discrete Cosine Transform (DCT) code which has the MDS property, and the second part is the residual noise arising from the non-polynomial nature of the target function. By using this decomposition, we show that popular DCT decoding algorithms customized to work in precision noise environments can be used to nullify the erroneous computations introduced by some worker nodes with high probability (see Section \ref{sec:err correction SBACC}). Towards characterizing the accuracy of SBACC, first, we prove that the approximation error of SBACC is approximately same as that of the vanilla BACC in the absence of untrusted workers (see Section \ref{sec:straggler resilent new scheme}). However, with a few workers that return erroneous computations, we again derive upper bounds on the accuracy of SBACC and show that it provides a choice in selecting the dimension of the DCT code as well as the number of evaluation points for Berrut's interpolation. As the main takeaway, we use experimental results to show that the computational accuracy of SBACC is significantly higher than that of BACC, and marginally higher than that of ApproxIFER (see Section \ref{sec:secure BACC s=0}). 
\section{Secure berrut approximated coded Computing}
\label{sec:SBACC}

The SBACC setup consists of a master server, which is connected to $N$ workers, denoted by $\mathcal{W}_{0}, \mathcal{W}_{1},\ldots, \mathcal{W}_{N-1}$ via dedicated links. Along the lines of BACC in \cite{b7}, we assume the presence of $S$ stragglers among the worker nodes, for some $1 \leq S  < N-2$. However, in contrast to \cite{b7}, we consider a setting wherein $A$ out of the remaining $M = N-S$ non-straggler worker nodes may send erroneous computation to the master server. Henceforth, throughout the paper, these $A$ worker nodes are referred to as adversaries.

Let $\mathbf{X}=(\mathbf{X}_0,\ldots,\mathbf{X}_{K-1})$ be the dataset at the master where $\mathbf{X}_j \in \mathbb{R}^{m\times n}$ for $j \in [K]$ such that $[K] \triangleq \{0, 1,2, \ldots,K-1\}$. The main objective of SBACC is to approximately evaluate “an arbitrary function” $f:\mathbb{R}^{m\times n} \rightarrow \mathbb{R}^{m\times n}$ over the matrices in $\mathbf{X}=(\mathbf{X}_0,\ldots,\mathbf{X}_{K-1})$ in a decentralized fashion while guaranteeing resiliency against $S$ stragglers and $A$  adversaries, respectively. More specifically, given that $f(\cdot)$ can be a non-polynomial function, the aim of SBACC is to ensure distributed computation of the function $f(\cdot)$ on $\mathbf{X}_{j}$ for $j\in [K-1]$ in a numerically stable manner with bounded errors, i.e., if $\mathbf{Y}_{j}$ denotes the result of distributed computation of the function $f(\cdot)$ on $\mathbf{X}_{j}$ then we need $\mathbf{Y}_{j} \approx f(\mathbf{X}_{j})$  for $j\in [K-1]$ with tolerable accuracy loss. In the following subsections, we will provide further details by describing various steps of distributed computation.

\subsection{Encoding in SBACC}
\label{sec:Encoding SBACC}

This section outlines a scheme to encode the data set $\mathbf{X}$ into shares to distribute them among the $N$ workers. Using Berrut's rational interpolant \cite{b7}, the master constructs a rational function $u(z)$ in the indeterminate $z$, as
\[
u(z) = 
\sum_{j=0}^{K-1} 
\frac{
\frac{(-1)^j}{z -\alpha_{j}}
}{
\sum_{k=0}^{K-1} \frac{(-1)^k}{z - \alpha_{k}}
}\mathbf{X}_{j},
\]
where $\alpha_j$'s are Chebyshev points of the first kind, defined as,
$\alpha_j = \cos \left( \frac{(2j + 1)\pi}{2K} \right)$. Note that, by the definition of Berrut's interpolant, we have $u(\alpha_{j})= \mathbf{X}_{j}$, for $j\in [K-1].$ Upon constructing $u(z)$, the master computes its evaluation at $z_{i} \in \mathbb{R}$, and shares $\mathbf{U}_{i} = u(z_{i})$ with the worker $\mathcal{W}_{i}$ for $i \in [N]$. In this work, we propose to choose $z_{i}$'s from Chebyshev points of the first kind, indicated by
\begin{equation}
\label{cheb_first_kind}
z_i = \cos \left( \frac{(2i + 1)\pi}{2N} \right),\quad i\in [N-1].
\end{equation}
\begin{remark}
Unlike in \eqref{cheb_first_kind}, in the straggler-resilient BACC scheme \cite{b7}, $z_{i}$'s are chosen from Chebyshev points of the second kind, indicated by $z_i = \cos \left( \frac{i \pi}{N} \right)$, for $i \in [N-1]$.
\end{remark}
\subsection{Computation at the Workers}
\label{sec:computation at worker}
After receiving its share $\mathbf{U}_{i}$ from the master, the worker $\mathcal{W}_{i}$ intends to compute $\mathbf{V}_i= f(\mathbf{U}_{i})$ for $i \in [N]$ and return the same to the master. Among the workers let $\mathcal{F} \subset [N]$ denote the indices of the stragglers that do not return their computations to the master. As a result, we will only focus on the computations at the non-straggler workers with indices $[N]/\mathcal{F}$ such that, $\lvert [N]/\mathcal{F} \rvert = M$. In addition, as the SBACC scheme accounts for the presence of a few adversarial workers, let $i_{1}, i_{2}, \ldots, i_{A} \in [N]/\mathcal{F}$ be their indices, and their computations be denoted by $\mathbf{V}_{i_{a}} +  \mathbf{E}_{i_{a}}$, for $i_{a} \in[N]/\mathcal{F}$. Here, $\mathbf{E}_{i_{a}} \in \mathbb{R}^{m \times n}$ represents the noise matrix injected by the adversarial worker $\mathcal{W}_{i_{a}}$. Furthermore, owing to the precision noise added due to floating point operations at the workers, the end results returned to the master are denoted by 
\begin{equation*}
    \mathbf{R}_{i} = \mathbf{V}_i + \mathbf{E}_{i} + \mathbf{P}_{i} \in \mathbb{R}^{m \times n}, i \in \{i_{1}, i_{2}, \ldots, i_{A}\},
\end{equation*}
\begin{equation*}
    \mathbf{R}_{i} =  \mathbf{V}_i + \mathbf{P}_{i} \in \mathbb{R}^{m \times n}, i \notin \{i_{1}, i_{2}, \ldots, i_{A}\} \cup \mathcal{F},
\end{equation*}
where $\mathbf{P}_{i} \in \mathbb{R}^{m\times n}$ captures the precision noise introduced by the worker $\mathcal{W}_{i}$. We assume that $\{\mathbf{P}_{i}\}$ are statistically independent across $i$, and the components of each $\{\mathbf{P}_{i}\}$ are independent and identically distributed as $\mathcal{N}(0, \sigma_{P}^{2})$.


\subsection{Error Correction in SBACC}
\label{sec:err correction SBACC}

The master collects the computations from the set of non-straggling worker nodes, whose indices are given by $[N]/\mathcal{F} = \{l_{1}, l_{2}, \ldots, l_{M}\}$. In particular, the corresponding set of computations are denoted by $\{\mathbf{R}_{l_{1}}, ~\mathbf{R}_{l_{2}}, ~\ldots~, \mathbf{R}_{l_{M}} \}$. Note that, for every $g \in {1, 2, \ldots, m}, h \in {1, 2, \ldots, n}$, let $\mathbf{r}_{g,h} = [\mathbf{R}_{l_{1}}(g, h) ~\mathbf{R}_{l_{2}}(g,h) ~\ldots~ \mathbf{R}_{l_{M}}(g,h)]$, represent a vector constructed from the $(g,h)$-th entry of each $\mathbf{R}_{l_{i}}$ for $l_{i} \in [N]/\mathcal{F}$. Similarly, let $\mathbf{v}_{g,h}$, $\mathbf{e}_{g, h}$ and $\mathbf{p}_{g,h}$ represent the corresponding $M$-length vectors carved from the $(g,h)$-th entry of $\{\mathbf{V}_{l_{i}}\}$, $\{\mathbf{E}_{l_{i}}\}$ and $\{\mathbf{P}_{l_{i}}\}$, respectively. 

Given that $\mathbf{r}_{g,h}$ is a noisy version of $\mathbf{v}_{g,h}$, the master intends to nullify the erroneous computations inserted by the adversaries. Towards that direction, we revisit some basic properties of Discrete Cosine Transform (DCT). Let $\Theta \in \mathbb{R}^{N \times N}$ represent the DCT matrix whose $(\theta_{1}, \theta_{2})$-th element is represented as $\Theta(\theta_{1}, \theta_{2}) = \sqrt{\frac{2}{N}} \, \beta(\theta_{1}) \cos\left(\frac{(2\theta_{2} + 1)\theta_{1}\pi}{2N}\right)$, for $0\leq \theta_{1}, \theta_{2} \leq N - 1,$ where, $\beta(0) = \frac{1}{\sqrt{2}}, \beta(\theta_{1}) = 1,\theta_{1} \neq 0$. 

\begin{proposition}
The following results with respect to the DCT matrix $\Theta$ are well known \cite{b9}, \cite{b10}, \cite{b11}. For any $K_{1}$ such that $1 \leq K_{1} < N$,
\begin{itemize}
    \item the first $K_{1}$ rows of the DCT matrix, denoted by $\mathbf{G} \in \mathbb{R}^{K_{1} \times N}$, can be used as a generator matrix of a real linear code. Such an $(N, K_{1})$ code is referred to as the DCT code
    \item the remaining $N-K_{1}$ rows of the DCT matrix, denoted by $\mathbf{H} \in \mathbb{R}^{N-K_{1} \times N}$, serve as the parity check matrix of the DCT code.
    \item DCT codes satisfy the MDS property, akin to Reed-Solomon codes in the parallel world of finite fields.
\end{itemize}
\end{proposition}
Using the above properties, we present our first result.
\begin{proposition}
For any $K_{1} \in \mathbb{N}$ such that $1< K_{1} < M$, the vector $\mathbf{r}_{g,h} \in \mathbb{R}^{M}$, $\forall g, h$ can be represented as a noisy codeword of a $K_{1}$-dimensional Discrete Cosine Transform (DCT) code of blocklength $M$.
\end{proposition}
\begin{IEEEproof}
First, we present the proof for $M = N$, and subsequently generalize it to any $M < N$. Suppose that $f(u(\cdot))$ is infinitely differentiable, such that it has Taylor series expansion at its evaluations $\{u(z_{i}), i \in [N]\}$. In that case, the components of the vector $\mathbf{v}_{g,h}$ have Taylor series expansion, and therefore, $\mathbf{v}_{g,h}$ can be written as 
\begin{equation}
    \label{eq:taylor}
    \mathbf{v}_{g,h} = \mathbf{t}_{g,h} + \mathbf{q}_{g,h}, 
\end{equation}
where $\mathbf{t}_{g,h}$ denotes the sum of the first $K_{1}$ term of the Taylor series, for some $K_{1} \in \mathbb{N}$ and $\mathbf{q}_{g, h}$ denotes the sum of the higher-order terms of the expansion. In the rest of the proof, we show that $\mathbf{t}_{g,h}$ is a codeword of a $(N, K_{1})$ DCT code, which in turn implies that $\mathbf{v}_{g,h}$ is a noisy DCT codeword due to \eqref{eq:taylor}, and so is $\mathbf{r}_{g,h}$ since it is an additive noisy version of $\mathbf{t}_{g,h}$. Henceforth, suppose that $\mathbf{t}_{g,h}$ is represented as an $N$-length row vector. Since the evaluation points $\{z_{i}, i \in [N]\}$ are the Chebyshev points of first kind, we can write $\mathbf{t}_{g,h}$ as $\mathbf{t}_{g,h} = \mathbf{m}\mathbf{Y}$, where  

\begin{small}
\begin{equation}
\label{eq:generator matrix}
\mathbf{Y} = \begin{bmatrix}
  1  &  1  &  \cdots & 1 \\
\vspace{0.25cm}
\cos \frac{\pi}{2N} & \cos \frac{3\pi}{2N} & \cdots & \cos \frac{(2N-1)\pi}{2N} \\

\cos^{2}\frac{\pi}{2N} & \cos^{2} \frac{3\pi}{2N} &  \cdots & \cos^{2} \frac{(2N-1)\pi}{2N} \\

\vdots & \vdots & \vdots  & \vdots \\

\cos^{K_{1}-1} \frac{\pi}{2N} & \cos^{K_{1}-1} \frac{3\pi}{2N} & \cdots & \cos^{K_{1}-1} \frac{(2N-1)\pi}{2N}
\end{bmatrix}
\end{equation}
\end{small}
\noindent is a $K_{1} \times N$ Vandermonde matrix, and $\mathbf{m} \in \mathbb{R}^{1 \times K_{1}}$ is a vector representing the coefficients of the first $K_{1}$ terms of the Taylor series expansion. Furthermore, from \cite{b9}, it is well known that the generator matrix $\mathbf{G}$ of a DCT code and the Vandermonde matrix $\mathbf{Y}$ in \eqref{eq:generator matrix} satisfy the relation $\mathbf{G} = \mathbf{B}\mathbf{Y}\mathbf{Z}$, where $\mathbf{B}$ is a $K_{1} \times K_{1}$ full-rank lower triangular matrix and $\mathbf{Z}$ is an $N \times N$ full-rank diagonal matrix. Using the above decomposition, we can further represent $\mathbf{t}_{g,h} = \mathbf{\bar{m}}\mathbf{B}\mathbf{Y}$, 
where $\mathbf{\bar{m}} = \mathbf{m} \mathbf{B^{-1}}$. Since $\mathbf{G} = \mathbf{B}\mathbf{Y}\mathbf{Z}$, the vector $\mathbf{\bar{m}}\mathbf{B}\mathbf{Y}$ is indeed a codeword of DCT code rotated by the diagonal matrix $\mathbf{Z}$. Therefore, the distance properties of the linear code with generator matrix $\mathbf{B}\mathbf{Y}$ remains identical to that of the linear code generated by $\mathbf{G}$. Therefore, when $M = N$, the $\mathbf{r}_{g,h}$ represents a corrupted codeword of an $(N,K_{1})$ DCT code. In general, when $M < N$, $\mathbf{t}_{g,h}$ can be seen as a punctured codeword of a DCT codeword of block-length $N$. This completes the proof.
\end{IEEEproof}
\mbox{}
\vspace*{0.01in}
Given that $\mathbf{r}_{g, h}$ is a noisy DCT codeword, the following result shows that the erroneous computations returned by the adversaries can be compensated under some conditions.

\begin{proposition}
\label{prop:dct err corr}
Let the SBACC setting be such that the composite function $f(u(\cdot))$ is a polynomial of degree $K_{1} - 1$. Under such a scenario, the erroneous computations returned by the $A$ adversarial workers can be accurately nullified as long as $A \leq \lfloor \frac{M-K_{1}}{2}\rfloor$ and the precision noise is zero.
\end{proposition}
\begin{IEEEproof}
Since the Taylor series on $f(u(\cdot))$ is bounded and the precision noise is zero, we have $\mathbf{r}_{g,h} = \mathbf{v}_{g, h} + \mathbf{e}_{g, h}$ such that $\mathbf{v}_{g, h} = \mathbf{t}_{g,h}$ and, $\mathbf{t}_{g,h}$ is a DCT codeword. Furthermore, the Hamming weight of $\mathbf{e}_{g, h}$ is $A$. For the generator matrix $\mathbf{G}$ of the DCT code, the corresponding parity check matrix $\mathbf{H}$ can also be written as $\mathbf{H} = \mathbf{A}\mathbf{T}\mathbf{W}$ such that $\mathbf{T}$ is an $N-K_{1} \times N$ Vandermonde matrix, $\mathbf{A}$ is an $N-K_{1} \times N-K_{1}$ full-rank lower triangular matrix and $\mathbf{W}$ is an $N \times N$ full-rank diagonal matrix. As a result, the modified parity check matrix $\mathbf{T}\mathbf{W}$ can be used on $\mathbf{r}_{g,h}$ to obtain the syndrome vector, and subsequently apply the BCH decoding algorithm for error localization and error correction in the presence of adversaries. Given that DCT code has the MDS property \cite{b9},\cite{b11}, this implies that the error vector $\mathbf{e}_{g,h}$ of Hamming weight $A$ can be accurately nullified as long as $A \leq \lfloor \frac{M-K_{1}}{2}\rfloor$.
\end{IEEEproof}

In the context of this work, the composite function $f(u(\cdot))$ is not a finite degree polynomial, and moreover, $\mathbf{r}_{g,h}$ is perturbed by precision noise along with the noise added by the adversaries. Under such a scenario, for any $K_{1}$, such that $1< K_{1} < M$, the received vector $\mathbf{r}_{g,h}$ can be treated as a $K_{1}$-dimensional noisy DCT codeword. Subsequently, a suitable syndrome based BCH decoding method can be applied to nullify the error vector $\mathbf{e}_{g,h}$. More specifically, to implement these ideas in precision noise scenarios, various coding theoretic-based and subspaced-based approaches are available \cite{b12},\cite{b13},\cite{b14},\cite{b15},\cite{b16}, \cite{b17}. Their implementation involves: (i) computation of the syndrome vector, (ii) estimation of the number of errors, (iii) identifying the location of errors, and (iv) estimation of error.

Upon applying the DCT based error-correction algorithm on the received computations $\{\mathbf{R}_{l_{1}}, ~\mathbf{R}_{l_{2}}, ~\ldots~, \mathbf{R}_{l_{M}} \}$, the master recovers $\{\mathbf{C}_{l_{1}}, ~\mathbf{C}_{l_{2}}, ~\ldots~, \mathbf{C}_{l_{M}} \}$, where 
$\mathbf{C}_{l_{i}} = \mathbf{R}_{l_{i}} - \hat{\mathbf{E}}_{l_{i}}, \forall i$ such that $\hat{\mathbf{E}}_{l_{i}}$ is the estimate of $\mathbf{E}_{l_{i}}$ introduced by error correction mechanism. Thus, we have 
\begin{equation}
\label{eq:rec_code1}
\mathbf{C}_{i} = \mathbf{V}_i  + \mathbf{P}_{i} + \mathbf{E}_{i} - \hat{\mathbf{E}}_{i}, i \in \{i_{1}, i_{2}, \ldots, i_{A}\},
\end{equation}
\begin{equation}
\label{eq:rec_code2}
\mathbf{C}_{i} =  \mathbf{V}_i + \mathbf{P}_{i} - \hat{\mathbf{E}}_{i}, i \notin \{i_{1}, i_{2}, \ldots, i_{A}\} \cup \mathcal{F}.
\end{equation}
\begin{remark}
The choice of $K_{1}$ determines the error correcting capability of the DCT code. As a consequence, it affects the accuracy with which the DCT decoder can correct the errors.
\end{remark}

\subsection{Function Reconstruction in SBACC}
\label{sec:reconstruction SBACC}
In the reconstruction process, the master uses $\{\mathbf{C}_{l_{i}}\}_{i \in [M]}$ and the Berrut's interpolants to obtain an approximation given by
\begin{equation}
\label{eq:berrut_reconstruction}
r_{\text{Berrut}, f}(z) = 
\sum_{r=0}^{M-1} \frac{\frac{(-1)^r}{z - \bar{{z}}_{r}} }{
\sum_{k=0}^{M} \frac{(-1)^k}{z - \bar{z}_{k}}} ~ \mathbf{C}_{l_{r+1}} ,
\end{equation}
where $\bar{z}_{r} = z_{l_{r+1}}$, such that $z_{l_{r+1}} = \mbox{cos}\left( \frac{(2l_{r+1} + 1)\pi}{2N} \right)$. Finally, to recover $\mathbf{Y}_{j}$, for $j \in [K-1]$, the master obtains an approximate version of $f(\mathbf{X}_{j})$ as $\mathbf{Y}_{j} = r_{\text{Berrut}, f}(\alpha_{j})$, for $j \in [K-1]$.


Similar to the existing BACC scheme, there is no requirement on the minimum number of worker nodes to return their computations in our framework. For a given set of stragglers among the worker nodes, the average accuracy of SBACC is
\begin{equation}
\label{eq:berrut approximation error}
\mathbb{E}_{\{\mathbf{P}_{i}\}, \{\mathbf{E}_{i}\}}||r_{\text{Berrut}, f}(z) - f(u(z))||,
\end{equation}
where $r_{\text{Berrut}, f}(z)$ is given in \eqref{eq:berrut_reconstruction} and the expectation operation is over the distribution of the precision noise matrices $\{\mathbf{P}_{i}, i \in [N-1]\}$ and adversarial noise $\{\mathbf{E}_{i}\}$. Note that \eqref{eq:berrut approximation error} depends on $M$, $A$, $\sigma_{P}^{2}$, and $K_{1}$. 

\section{Straggler-resilience of SBACC Scheme}
\label{sec:straggler resilent new scheme}
Before evaluating the approximation error of SBACC in \eqref{eq:berrut approximation error}, we study its special case by considering only the presence of stragglers, i.e., with $A = 0$ and $\sigma_{P}^{2} = 0$. As a result, \eqref{eq:berrut approximation error} collapses to a deterministic quantity $||r_{\text{Berrut}, f}(z) - f(u(z))||$. Since Chebyshev points of the first kind are used in SBACC, the upper bounds on the approximation error of BACC is not applicable in this context. Before presenting an upper bound on $||r_{\text{Berrut}, f}(z) - f(u(z))||$ for our case, we give an upper bound on the Lebesgue constant when using the Chebyshev points of the first kind.

\begin{lemma}
\label{lemma:lebesgue}
 Let $\mathcal{X}_{M}$ be a set with $M$ ordered, distinct interpolation points chosen from a set $\mathcal{X}_N\!=\!\left\{x_{i}\right\}_{i=0}^{N-1}$, where, $M=N-S$, and $x_{i}\!=\!\cos \left( \frac{(2i + 1)\pi}{2N} \right), i \in [N-1]$. Then, $\mathcal{X}_{M}$ are well-spaced points for $2\leq M < N$. Therefore, the Lebesgue constant for Berrut's rational interpolant in $\mathcal{X}_{M}$ is upper bounded as,
 \begin{eqnarray*}
 \Lambda _n\leq \left (\frac{(S+1)(S+4)\pi ^2}{8}+1\right)\big (1+\pi ^2(S+1)\ln (N-S)\big). 
 \end{eqnarray*}
\end{lemma}

Lemma \ref{lemma:lebesgue} shows that, similar to the straggler resilient  BACC scheme in \cite{b7}, the Lebesgue constant for the proposed SBACC scheme varies as $\ln (N-S)$.

\begin{theorem}
\label{Th:theorem 1}
Let $r_{\text{Berrut},f}(z)$ be defined by \eqref{eq:berrut_reconstruction} and $g(z) = f(u(z))$ have a continuous second derivative on $[-1,1]$. For SBACC with $N$ worker nodes and $S$ stragglers, where $S<N-2$, when $A=0$ and $\sigma_{P}^2 = 0$ the approximation error defined in \eqref{eq:berrut approximation error} is upper bounded as,
\begin{eqnarray*}
\left\|r_{\text {Berrut},f}(z)-g(z)\right\| \leq 2\Delta(1+R) \sin \left(\frac{(S+1) \pi}{2 N}\right),
\end{eqnarray*}
where, $\Delta=\left\|g^{\prime \prime}(z)\right\|$, if $N-S$ is odd and $\Delta=\left(\left\|g^{\prime \prime}(z)\right\|+\left\|g^{\prime}(z)\right\|\right)$, when $N-S$ is even and $R=\frac{(S+1)(S+4) \pi^2}{8}$.
\end{theorem}

 \begin{figure*}
      \begin{scriptsize}
\begin{IEEEeqnarray}{rCl}\label{eq:adv_bound}
&&\mathbb{E}_{\{\mathbf{P}_{i}\}, \{\mathbf{E}_{i}\}}\left\|r_{\text{Berrut}, f}(z) \!- \! f(u(z))\right\|^2 
\leq \left|\left|2\Delta(1+R) \sin\left( \frac{((N-N_{1})+1)\pi}{2N} \right)\! \right|\right|^2\! \!+\!\!N_{1} \sigma_p^2 \max_{r\in[N]} \frac{\left|\prod_{i\neq r}(z-z_i)\right|^2}{\left|\sum_{k=0}^{N_{1}-1}(-1)^k \prod_{i\neq k}(z-z_i)\right|^2}  \nonumber\\
&&+  2\sigma_q\sigma_A + (1-\mbox{Prob}(E_{Loc}))A\sigma_q^2 \max_{r\in[N]} \frac{\frac{1}{\left|z-\bar{z}_r\right|^2}}{\left|\sum_{k=0}^{N_{1}-1}\frac{(-1)^k}{z-\bar{z}_k}\right|^2}
+ \mbox{Prob}(E_{Loc}) \frac{N_{1}(N_{1}-1)\cdots(N_{1}-A+1)}{N(N-1)\cdots(N-A+1)} 
 2A\sigma_A^2 \max_{r\in[N]} \frac{\frac{1}{\left|z-\bar{z}_r\right|^2}}{\left|\sum_{k=0}^{N_{1}-1}\frac{(-1)^k}{z-\bar{z}_k}\right|^2} 
\end{IEEEeqnarray}
\end{scriptsize}
\hrule
 \end{figure*}
\subsection{Experimental Results}
\begin{figure}
\centering
\includegraphics[scale = 0.23]{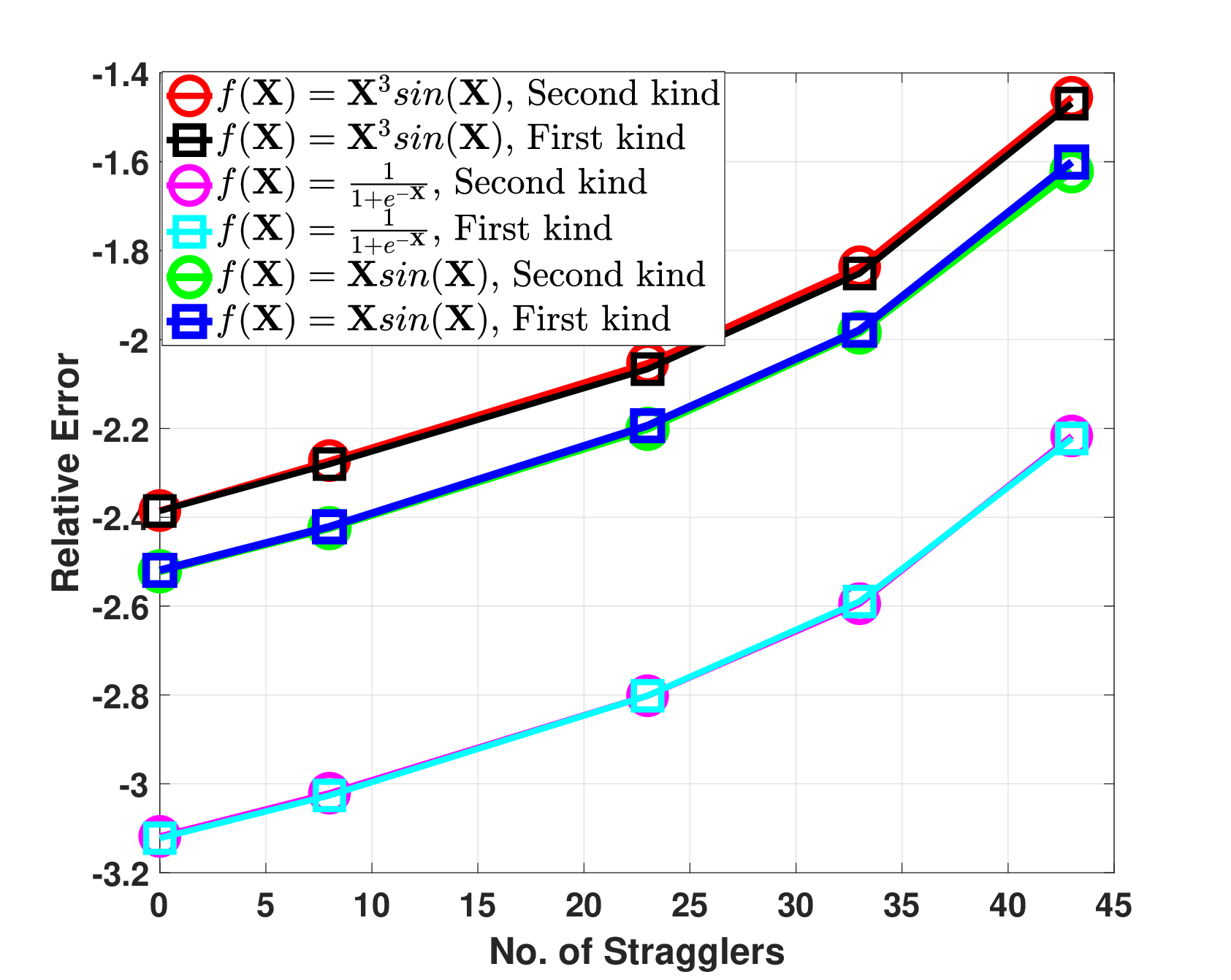}
\vspace{-0.1cm}
\caption{Average relative error of the approximation (in dB scale) of BACC and SBACC for various functions with parameters $N=53$, $K=4$, $A=0$, $\sigma_{P}^{2} = 0$.}
\label{fig: straggler vs accuracy}
\end{figure}

Lemma \ref{lemma:lebesgue} and Theorem \ref{Th:theorem 1} confirm that the bounds on the approximation error of SBACC is almost same as that for BACC, when $A = 0$ and $\sigma_{P}^{2} = 0$. To demonstrate this result, we implement SBACC and BACC to compute several non-polynomial functions. Here, $\mathbf{X}_{i} \in \mathbb{R}^{5\times 5}$ for $i \in [K]$, whose samples are drawn from uniform distribution. Assuming $\mathbf{Y}_{j} \approx f(\mathbf{X}_{j})$ to denote the approximate computation using BACC or SBACC for $j \in [K-1]$, and $\mathbf{Y}=f(\mathbf{X}_{j})$ to denote the centralized computation at the master without using BACC or SBACC. We define the relative error introduced by BACC and SBACC with respect to the centralized computation as $e_{rel} \triangleq \frac{||\mathbf{Y}-\mathbf{Y'}||^{2}}{||\mathbf{Y}||^{2}},$ and finally compute the average relative error for the approximate computation, and present them in Fig.\ref{fig: straggler vs accuracy} by varying $S$. The plots confirm that SBACC and BACC exhibit similar resiliency against stragglers.  

\section{Accuracy of SBACC with Adversaries}
\label{sec:secure BACC s=0}
In this section, we study the behavior of the average accuracy in \eqref{eq:berrut approximation error} by considering $A > 0$, $\sigma_{P}^{2} > 0$ and study its relation as a function of $K_{1}$ and the number of interpolation points for reconstruction, denoted by $N_{1}$ such that $2 \leq N_{1} \leq N$. For simpler description, we assume $S = 0$, although these results can be easily generalized when $S > 0$. Applying \eqref{eq:rec_code1} and \eqref{eq:rec_code2} in \eqref{eq:berrut_reconstruction}, when $S = 0$, the expression in \eqref{eq:berrut approximation error} becomes
\begin{equation}
\label{eq:err_Adv}
\mathbb{E}_{\{\mathbf{P}_{i}\}, \{\mathbf{E}_{i}\}}||r_{\text{Berrut},v}(z) + r_{\text{Berrut},p}(z) + r_{\text{Berrut},e}(z) - f(u(z))||
\end{equation}
where $r_{\text{Berrut},v}(z)$, $r_{\text{Berrut},p}(z)$ and $r_{\text{Berrut},e}(z)$ are the portions of \eqref{eq:berrut_reconstruction} contributed by $\{\mathbf{V}_{i}\}$, $\{\mathbf{P}_{i}\}$ and $\{\mathbf{E}_{i} - \hat{\mathbf{E}}_{i}\}$, respectively. Furthermore, when $\sigma_{P}^2>0$, let,  $\mbox{Prob}(E_{Loc})$ and $1 - \mbox{Prob}(E_{Loc})$ denotes the probability of imperfect and perfect localization in the DCT decoder. Under this assumption, the following theorem provides an upper bound on the average squared approximation error of SBACC when the entries of $\{\mathbf{E}_{i}\}$ are statistically independent and distributed as $\mathcal{N}(0, \sigma^{2}_{A})$.

\begin{theorem}
 For the SBACC scheme with $N$ workers, when $S=0$, $A>0$, $\sigma_{P}^2>0$, $\sigma_{A}^2>0$ and $2\leq N_{1}\leq N$, \eqref{eq:err_Adv} is upper bounded as given in \eqref{eq:adv_bound}, where, $\Delta =\left\|g^{\prime \prime}(z)\right\|$, when $N_{1}$ is odd and $\Delta=\left(\left\|g^{\prime \prime}(z)\right\|+\left\|g^{\prime}(z)\right\|\right)$, when $N_{1}$ is even.
\end{theorem}

\vspace{-0.5cm}
\begin{figure}[ht!]
\centering
\includegraphics[scale = 0.29]{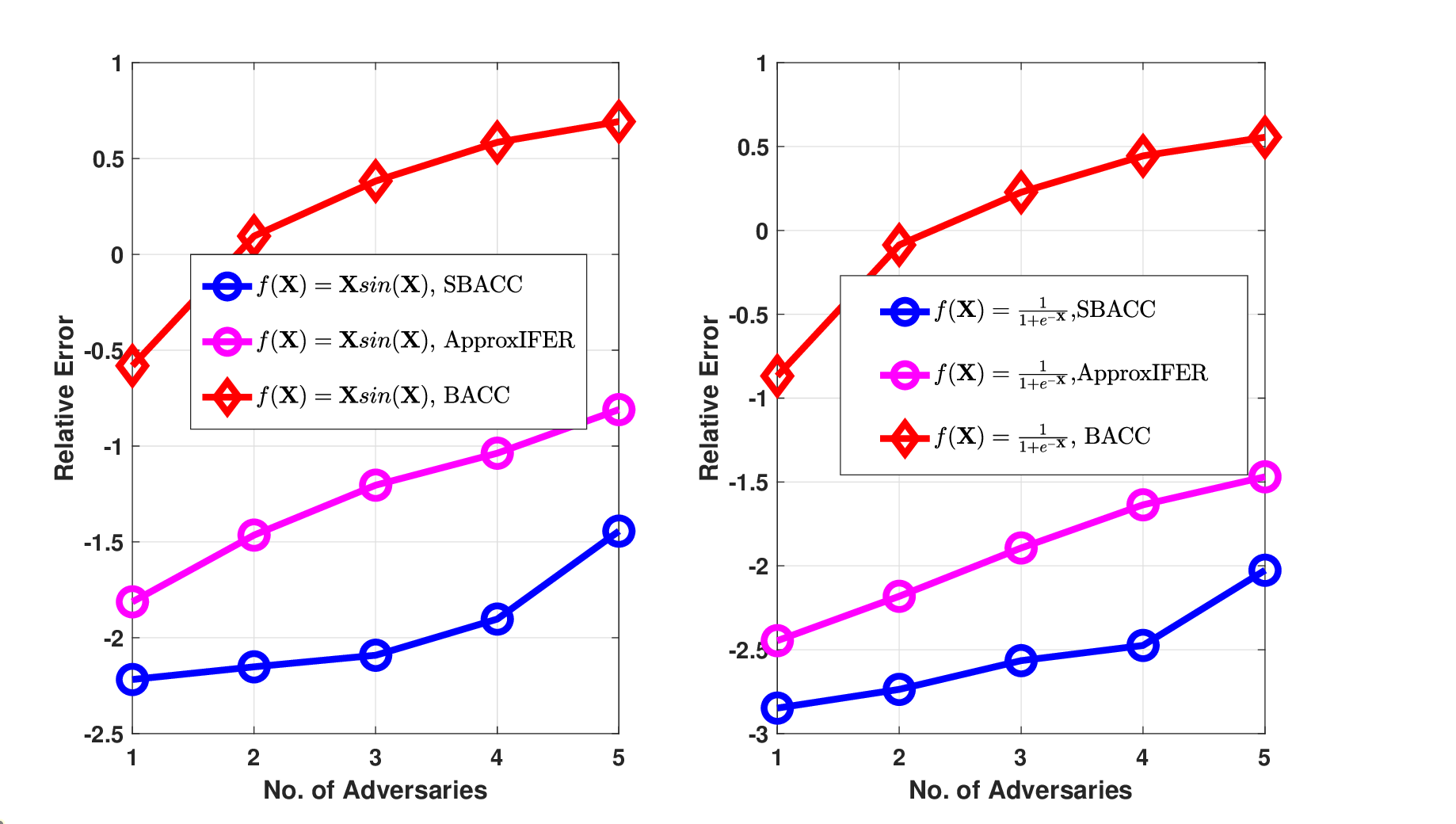}
\vspace{-0.7cm}
\caption{Average relative error (in dB scale) for BACC, ApproxIFER and SBACC  as a function of $A$ with $N=53$, $K=4$, $S=0$, $K_{1}=43$, $N_{1}=35$.  Here the entries of $\{\mathbf{E}_{i_{a}}\}$ are i.i.d. as $\mathcal{N}(0, 10^4)$.}
\label{fig: adv vs accuracy}
\end{figure}
To demonstrate the accuracy benefits of SBACC scheme over ApproxIFER and BACC we present experimental results on average relative error of all three schemes in Fig. \ref{fig: adv vs accuracy}. The plots confirm that the accuracy of BACC degrades as $A$ increases. However, SBACC shows significant improvement in accuracy over that of BACC and a remarkable improvement over that of ApproxIFER \cite{a3} because of its correction capability. Observe that the accuracy of SBACC also degrades as $A$ increases, however, this behavior is well known due to the error correction in floating point environment \cite{b9}. 

To implement SBACC in practice, \eqref{eq:adv_bound} can be used to choose $K_{1}$ and $N_{1}$ for a fixed $N$, $A > 0$, $\sigma_{A}^{2}>0$ and $\sigma_{P}^{2}>0$. To characterize the behavior of \eqref{eq:adv_bound}, as a function of $N_{1}$, for a fixed $K_{1}$, the first term of \eqref{eq:adv_bound} decreases when $N_{1}$ increases (by Theorem \ref{Th:theorem 1}), while the remaining terms increase. Due to this conflicting trends, we believe there exists an optimal value of $N_{1}$ that minimizes \eqref{eq:adv_bound}. Similarly, for a given $N_{1}$, the third and fourth terms are dependent on $K_{1}$ owing to the effect of residual noise during Taylor series approximation as well as the varying dimension of the parity check matrix. Thus, we believe there exists an optimal value of $K_{1}$ that minimizes \eqref{eq:adv_bound}. Stitching together, we believe that \eqref{eq:adv_bound} can be minimized to jointly obtain the optimal combination of $N_{1}$ and $K_{1}$.

$~~$\end{document}